\documentclass[secnumarabic,amssymb, nobibnotes, twocolumn,reprint, aip, ]{revtex4-1}

\usepackage{graphics}

\usepackage{graphicx}

\begin{document}

\title{Fiber-Cavity-Based Optomechanical Device}

\author{N. E. Flowers-Jacobs}
\email{nathan.flowers-jacobs@yale.edu}
\affiliation{Department of Physics,Yale University, New Haven, CT  06520, USA}
\author{S. W. Hoch}
\affiliation{Department of Physics,Yale University, New Haven, CT  06520, USA}
\author{J. C. Sankey}
\affiliation{Department of Physics, McGill University, Montreal, Quebec, Canada}
\author{A. Kashkanova}
\affiliation{Department of Physics,Yale University, New Haven, CT  06520, USA}
\author{A. M. Jayich}
\affiliation{Department of Physics,Yale University, New Haven, CT  06520, USA}
\author{C. Deutsch}
\affiliation{Laboratoire Kastler Brossel, ENS/UPMC-Paris 6/CNRS, F-75005 Paris, France}
\author{J. Reichel}
\affiliation{Laboratoire Kastler Brossel, ENS/UPMC-Paris 6/CNRS, F-75005 Paris, France}
\author{J. G. E. Harris}
\affiliation{Department of Physics,Yale University, New Haven, CT  06520, USA}
\affiliation{Department of Applied Physics,Yale University, New Haven, CT  06520, USA}
\date{\today}

\begin{abstract}
We describe an optomechanical device consisting of a fiber-based optical cavity containing a silicon nitiride membrane. In comparison with typical free-space cavities, the fiber-cavity's small mode size (10~$\mathrm{\mu m}$ waist, 80~$\mathrm{\mu m}$ length) allows the use of smaller, lighter membranes and increases the cavity-membrane linear coupling to 3~GHz/nm and the quadratic coupling to $\mathrm{20~GHz/nm^2}$.  This device is also intrinsically fiber-coupled and uses glass ferrules for passive alignment. These improvements will greatly simplify the use of optomechanical systems, particularly in cryogenic settings.  At room temperature, we expect these devices to be able to detect the shot noise of radiation pressure.
\end{abstract}

\maketitle

  In quantum mechanics, a measurement of one variable is accompanied by back-action on the conjugate variable. In the particular case of an optical displacement measurement, the quantum back-action is radiation pressure shot noise (RPSN) \cite{Caves1981,Clerk2010}, the Poissonian noise in the momentum transferred by reflecting photons. 

  When a high-finesse cavity is used to increase the sensitivity of the displacement measurement, the RPSN is also increased. The connection between increased cavity finesse, increased measurement sensitivity, and increased RPSN can be understood qualitatively by noting that the number of times each photon interacts with a mechanical element inside a cavity is approximately equal to the cavity finesse.

  In the field of optomechanics, floppy mechanical elements are integrated into high-finesse optical cavities in order to observe various quantum effects, including RPSN\cite{Kippenberg2008, Brooks2011}. The goal of observing RPSN is motivated by basic questions about quantum measurements, as well as by the fact that RPSN is expected to limit the performance of next-generation gravitational-wave observatories (though squeezed light can be used to mitigate the effect)\cite{Kimble2001, Schnabel2011}. 

  To date, RPSN has not been observed in solid-state optomechanical devices, largely because it has been obscured by the thermal Langevin force that produces Brownian motion\cite{Verlot2009}. While it has been proposed that correlation measurements can be used to distinguish RPSN in the presence of a much larger Langevin force\cite{Verlot2009, Borkje2010}, such a measurement would be simplified by increasing the RPSN relative to the thermal Langevin force. There is an additional motivation for increasing the RPSN: the optomechanical generation of squeezed light (e.g. for improving the performance of gravitational-wave observatories) requires a setup in which the RPSN dominates over the Langevin force\cite{Mancini1994, Fabre1994}.
  
  Increasing the effect of RPSN in comparison to thermal motion involves optimizing both the mechanical system and the optical cavity.  The Langevin force has a single-sided power spectral density $S_{\text{F}}^{\text{thermal}}=4 m \gamma_{\text{m}} k_{\text{B}} T$, and so is reduced by decreasing the mechanical element's full-width at half-maximum (FWHM) linewidth $\gamma_{\text{m}}$, effective mass $m$, and temperature $T$.  The RPSN force has a single-sided power spectral density $S_{\text{F}}^{\text{RPSN}}=8 \hbar^2 A^2 \bar{n} / \kappa$ when the laser is resonant with the cavity and the mechanical resonance frequency $\omega_{\text{m}}$ is much less than the cavity FWHM linewidth $\kappa$ (the ``bad cavity'' limit).  Thus, the RPSN force is increased by increasing the optomechanical coupling $A=d\omega_{\text{cav}}/dx$, increasing the average number of photons stored in the cavity $\bar{n}$, and decreasing $\kappa$.
  
\begin{figure}
\begin{center}
\includegraphics[width=3.4in]{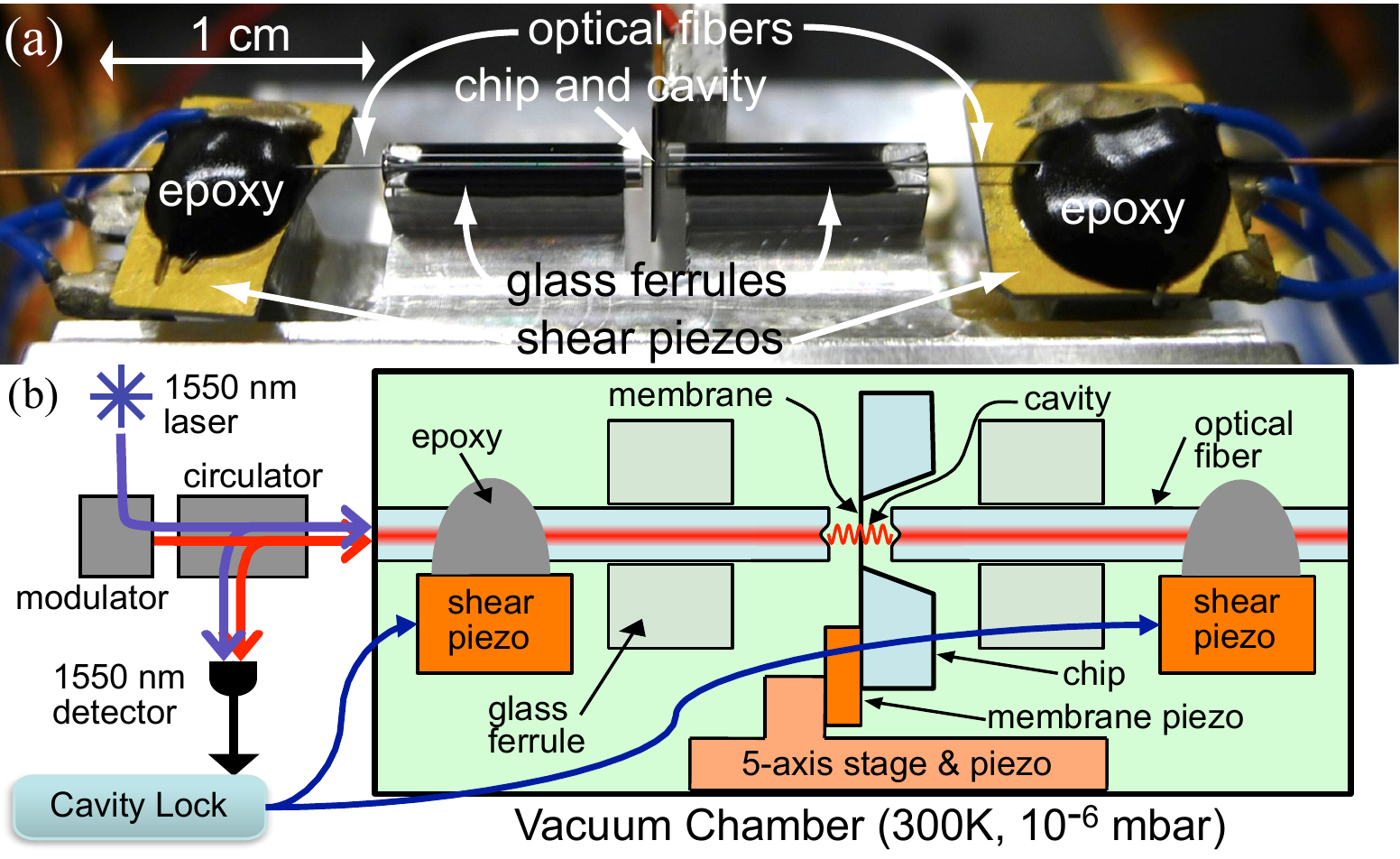}
\caption{(a) Photograph and (b) diagram of our experimental setup.  This optomechanical system dispersively couples a 80~$\mathrm{\mu m}$ long high-finesse $\mathcal{F}=15,000$ fiber-based optical cavity and a 150~nm thick, 250~$\mathrm{\mu m}$ square silicon nitride membrane.  The cavity is passively aligned using two glass ferrules.}
\label{fig:apparatus}
\end{center}
\end{figure}

  In this paper, we present a new type of ``membrane-in-the-middle'' optomechanical system (figure \ref{fig:apparatus}) that is designed to increase the RPSN force over the thermal Langevin force.  Instead of using the usual cm-scale free-space optical cavity \cite{Thompson2008, Sankey2010, WilsonRegal2009, Vitali2011}, we use an $\mathrm{80~\mu m}$ long cavity formed between the ends of two $\mathrm{200~\mu m}$ diameter single-mode optical fibers \cite{Metzger2004, Favero2009, Hunger2010}.  The end of each fiber is prepared using a $\mathrm{CO_{2}}$ laser to create a smooth concave depression with a $\mathtt{\sim}50\mathrm{~\mu m}$ diameter and a $\mathtt{\sim}300~\mathrm{\mu m}$ radius of curvature.  A mirror-coating with 99.99\% reflectivity at 1550~nm is then deposited on the fiber end by ATFilms \cite{ATFaddress}.

  To verify the quality of the mirror coatings, we measured the properties of a cavity formed between two of these fiber-mirrors \cite{Hunger2010}.  The two fiber-mirrors are aligned using a five-axis stage to form a $\mathrm{45~\mu m}$ long cavity.  After optimizing the alignment, we find that the coupling from the single-mode input fiber into the $\mathrm{TEM_{00}}$ (transverse electromagnetic) mode of the cavity is $\epsilon>50\%$ (the coupling $\epsilon$ is defined in the next paragraph) and the cavity has a finesse of $\mathcal{F}=27,000$.  Based on an independent measurement of the mirror-coating transmission $T=9\times10^{-5}$, light leaving the fiber-cavity is predominately transmitted through the mirror coating and is not absorbed or scattered.

  The input couplings given in this paper are estimates based on the ratio of on-resonance and off-resonance reflected power $R_{res}=P_{on}/P_{off}$.  Since the cavities discussed in this paper are double-sided with identical mirror coatings on both mirrors, we estimate the input mode coupling as $\epsilon=1-R_{res}$.  This estimate neglects two sources of loss which could lead to an error in the coupling estimate of as much as 20\%.  First, this estimate assumes that the cavity's internal loss is negligible compared to the light leaving the cavity through the mirrors.  Given the independent measurement of mirror transmission, this assumption is reasonable when the cavity finesse $\mathcal{F}\approx30,000$ but will underestimate the coupling when $\mathcal{F}<30,000$.  Second, angular misalignment of the mirror surface with the traveling mode in the fiber would reduce the fraction of off-resonant light reflected back into the fiber's traveling mode which would lead to an overestimate of the coupling.
 
  Although the translation stages allow us to optimize the alignment of the fiber mirrors (and thereby determine their optical properties), they do not provide a mechanically rigid platform for further measurements.  Instead we use custom fabricated glass ferrules (from Polymicro Technologies \cite{Polymicroaddress} with an inner diameter of $206\mathrm{\pm3~\mu m}$, an outer diameter of $1800\mathrm{\pm25\mu m}$, and an overall length of $10\mathrm{\pm0.5~mm}$) to passively align the fiber mirrors.  To a good approximation, the ferrules passively constrain the location of the two fibers in eight of their twelve degrees of freedom, leaving each fiber free to rotate in and translate along the ferrule.
  
  To test the passive alignment that can be achieved with these ferrules, we insert two fiber mirrors into a single glass ferrule to form a $\mathrm{25~\mu m}$ long cavity.  At this cavity length, we observe that the cavity's finesse and coupling are strongly dependent on the remaining two degrees of freedom (the rotation of each fiber relative to the ferrule). After rotating both fibers, we found a maximum finesse $\mathcal{F}=29,000$ and $\epsilon>60\%$, confirming that passive alignment does not degrade cavity performance.  The rotation dependence is presumably caused by the $\mathrm{\mu m}$-scale misalignment between the center of the concave depression and the center of the fiber \cite{Hunger2012}.

  The optomechanical device that is the central focus of this Letter is shown in figure \ref{fig:apparatus}.  In order to accommodate the mechanical oscillator, two identical ferrules are used which are separated by $\mathtt{\sim}1$~mm.  To achieve alignment, the two empty ferrules are first passively aligned to each other by inserting a $\mathrm{200~\mu m}$ diameter ``alignment'' glass fiber through both ferrules.  While this alignment fiber is in place, the ferrules are epoxied to an aluminum plate using Stycast 2850 epoxy (the black layer visible below the ferrules in figure \ref{fig:apparatus}a).  Once the epoxy is cured, the alignment fiber is removed.  The mirrored fibers are then inserted into the ferrules to form an $\mathrm{80~\mu m}$ long fiber-cavity at the center of the gap between the ferrules.  After rotating both fibers inside the ferrules, we achieve $\epsilon>40\%$ and $\mathcal{F}=19,000$.  This finesse is smaller than that of the test cavities described earlier, but the cavity is also significantly longer.  This is consistent with our observation that the finesse of the test cavities decreased as the cavity length increased and was generally $\mathcal{F}<20,000$ at a cavity length of $\mathrm{80~\mu m}$.
  
  To insert the membrane into the cavity, we partially retract the fibers and position the membrane between the ferrules using a vacuum-compatible five-axis alignment stage (Newport 9082-V).  The membrane is a 150~nm thick, 250~$\mathrm{\mu m}$ square of silicon nitride custom-fabricated by Norcada Inc \cite{Norcadaaddress}.  The membrane is formed by etching through a 200~$\mathrm{\mu m}$ thick, 5~mm square silicon chip.  The fundamental mode of this membrane has a resonance frequency of 1.74~MHz and a quality factor in vacuum of $Q=20,000$.  The $Q$ depends on the membrane's coupling to the mounting structure \cite{WilsonRegal2009}, and the comparatively low $Q$ of this membrane may reflect losses caused by epoxying the membrane chip to a piezo (see figure \ref{fig:apparatus}).  This piezo is used for fine control of the membrane's position along the cavity axis.  In future experiments $Q$ can be improved by using  membrane mounts that require less epoxy.
  
  After inserting the membrane, we reposition the fiber-mirrors to again form the $\mathrm{80~\mu m}$ cavity with the membrane in the middle (figure \ref{fig:apparatus}).  Because the chip is thicker than the cavity length, one of the fibers extends into the etch pit in the silicon chip.  With the cavity defined and the membrane in place, the fibers are attached to shear piezos by encapsulating the fiber in epoxy (black epoxy on top of the shear piezos' gold electrodes in figure \ref{fig:apparatus}a).  Fine-tuning of the position and orientation of the membrane is performed in air using the Newport stage to maximize the $\mathrm{TEM_{00}}$ mode's coupling and finesse \cite{Sankey2010}.  The experiment is then pumped to less than $\mathrm{10^{-6}~Torr}$.  In vacuum, we find that $\epsilon>50\%$ and $\mathcal{F}=15,000$.  This decrease in finesse does not have the dependence on membrane position that is observed when the membrane is scattering or absorbing light \cite{Jayich2008}, and is likely related to the position of the cavity mode on the end-mirrors and coupling to other TEM modes \cite{Sankey2010}.

\begin{figure}
\begin{center}
\includegraphics[width=3.4in]{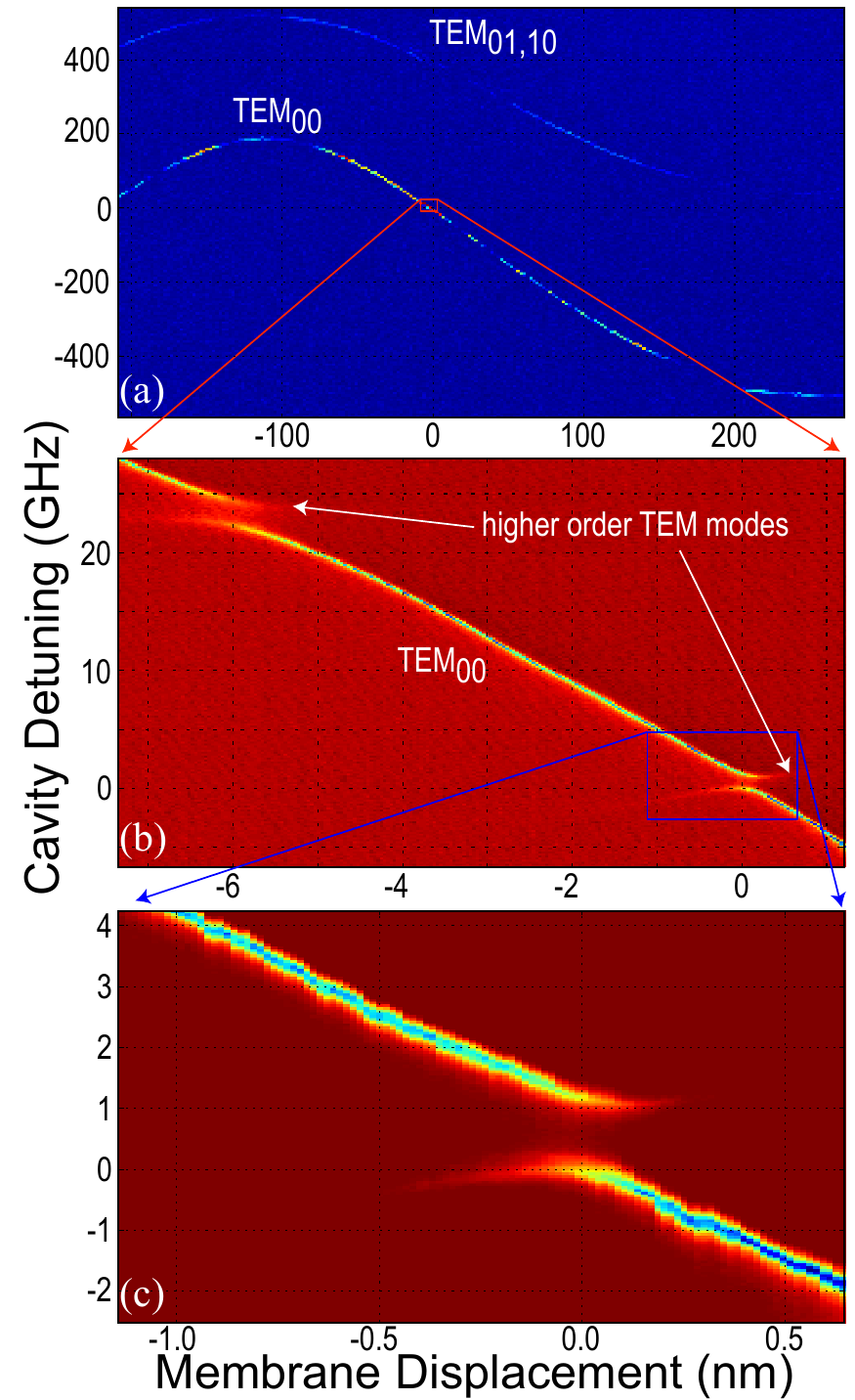}
\caption{ The transmission through the fiber-cavity (a) and reflection off of the fiber-cavity (b, c) as a function of cavity length (y-axis, converted into frequency units) and membrane position (x-axis).  The reflected signal (b, c) ranges from $R=1$ (off-resonance, dark red) to $R=0.5$ (on-resonance, dark blue).}
\end{center}
\label{fig:bandstructure}
\end{figure}
  
  To characterize the optomechanical properties of this device, we measure the cavity transmission and reflection versus cavity length and membrane position while the cavity is excited by a laser with wavelength $\lambda=1550$~nm.  An example of such a measurement is shown in figure 2\ref{fig:bandstructure}.  In figure 2\ref{fig:bandstructure}a the membrane's position is swept from a node of the electric field (at $-110$~nm) to an anti-node (at $280~\mathrm{nm}$) over 5~sec; simultaneously, the cavity length is oscillated with a 6~ms period and an amplitude of $\lambda/6$.  We observe both the $\mathrm{TEM_{00}}$ mode and, more faintly, the $\mathrm{TEM_{01,10}}$ modes.  We also observe significant changes in the magnitude of the resonant transmission as a function of membrane position.  As discussed below, these changes are caused by coupling to higher order cavity modes.
  
  In figure 2\ref{fig:bandstructure}b and 2\ref{fig:bandstructure}c we focus on an intersection between the $\mathrm{TEM_{00}}$ mode and a higher order mode.  The avoided crossings reflect the coupling between TEM modes due to a small misalignment of the membrane relative to the cavity waist \cite{Sankey2010}. The linear optomechanical coupling near this point is $A/2 \pi= 3$~GHz/nm and the avoided crossing has a quadratic coupling\cite{Sankey2010} of $\mathrm{20~GHz/nm^2}$.
    
  In order to measure the dynamics of the optomechanical system, we lock the cavity to the laser and use a heterodyne detection scheme to monitor the membrane's motion.  To achieve this, we modulate the laser to create a single sideband that is detuned from the carrier by 920~MHz and whose power is $\mathtt{\sim}10\%$ of the carrier (as shown in Figure \ref{fig:apparatus}b). The cavity length is tuned so that the smaller sideband is near the cavity resonance. As a result, the more intense carrier beam is promptly reflected from the cavity. A heterodyne technique is used to measure the magnitude and phase of the resulting beat note, thereby providing a measure of the detuning between the cavity and the laser sideband. Locking between the cavity and the laser is accomplished by using an FPGA (field-programmable gate array) to convert the inferred detuning into a voltage that is applied to the shear piezos that control the cavity length. The $\mathtt{\sim}5\mathrm{~kHz}$ bandwidth of this lock is much less than the membrane's 1.74~MHz fundamental resonance frequency.

\begin{figure}
\begin{center}
\includegraphics[width=3.4in]{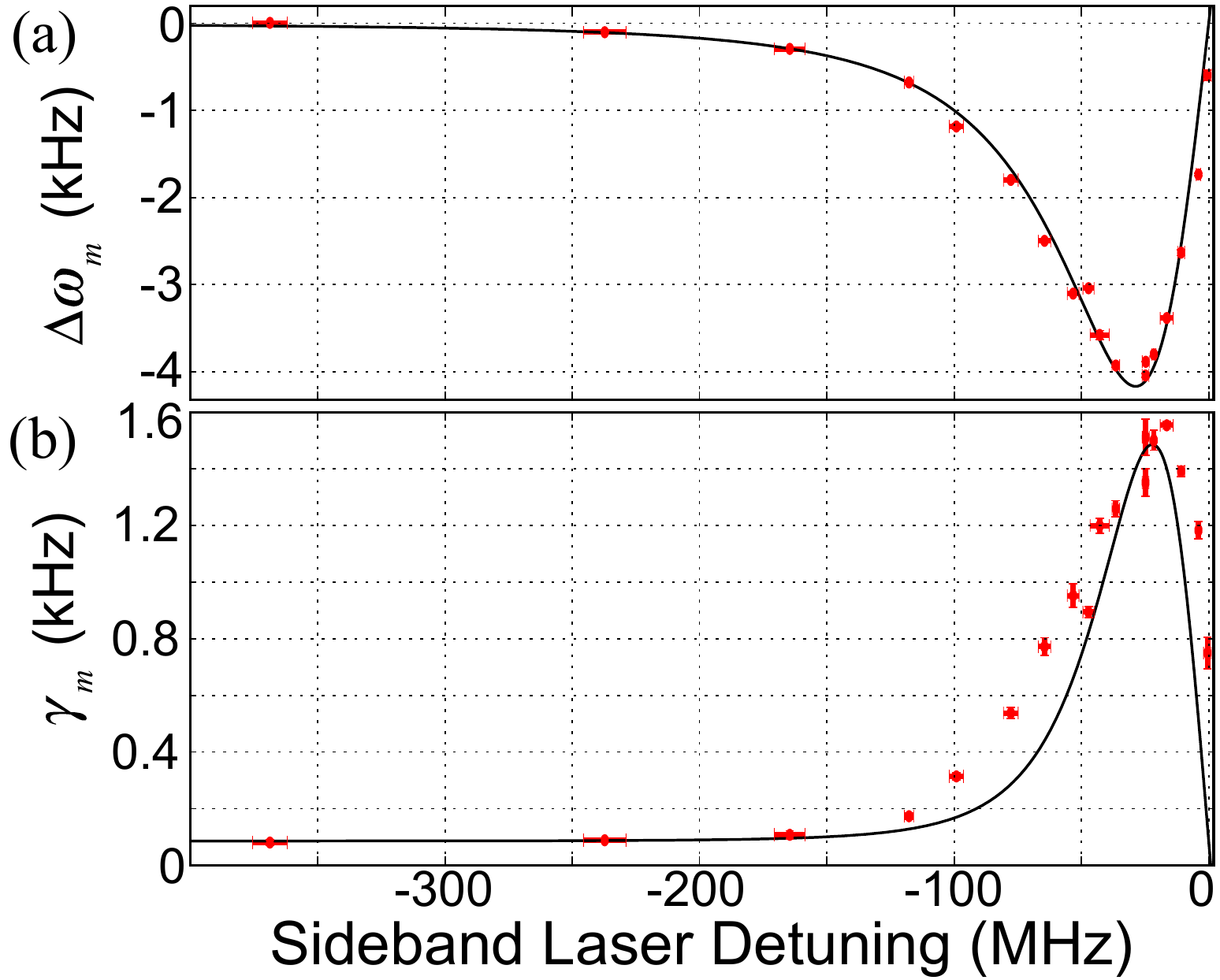}
\caption{We plot the measured shift $\Delta \omega_{\text{m}}$ in the mechanical resonance frequency $\omega_{\text{m}}$ from 5.2109~MHz (a) and the damping rate $\gamma_{\text{m}}$ (b) as a function of the detuning of the laser from the cavity for an incident mode-matched laser power of $\mathrm{8~\mu W}$ (points with statistical error bars and solid theory line).}
\end{center}
\label{fig:optomechanics}
\end{figure}

  We lock the cavity over a range of detunings and measure the membrane's Brownian motion by recording the reflected heterodyne signal. The Fourier transform of this record is fit to the expected form of a harmonic oscillator; this fit provides the mechanical damping $\gamma_{\text{m}}$ and resonance frequency $\omega_{\text{m}}$. We concentrate on a higher-order mechanical mode with $\omega_{\text{m}}/2\pi=\mathrm{5.2109~MHz}$ and $\gamma_{\text{m}}/2\pi=\mathrm{85~Hz}$ in the absence of optomechanical feedback.  Figure 3\ref{fig:optomechanics} shows a plot of $\gamma_{\text{m}}$ and $\omega_{\text{m}}$ a function of detuning (dots) and the theoretical expectation (lines) for cavity-induced radiation pressure feedback \cite{Marquardt2007}.  Since this device is in the ``bad cavity'' limit $\omega_{\text{m}}/2\pi\ll\kappa/2\pi=\mathrm{100~MHz}$, we observe the largest change in $\omega_{\text{m}}$ and $\gamma_{\text{m}}$ at a detuning $\mathtt{\sim}\kappa/4$ and the sign of the change is determined by the sign of the detuning.

  The observed change in $\omega_{\text{m}}$ agrees well with optomechanical theory, but fluctuations in detuning increase $\gamma_{\text{m}}$ in comparison to theory at detunings where $\omega_{\text{m}}$ is strongly dependent on detuning.  We also observe regenerative oscillations at positive detunings when the theoretical expectation for $\gamma_{\text{m}} \lesssim 0$;  this data is not included in figure 3\ref{fig:optomechanics}.
 
  This type of fiber-cavity has a number of practical advantages over standard free-space systems.  The cavity is inherently fiber-coupled, so it is simple to route light into the cavity and send a large fraction of the reflected and transmitted light to other parts of an experiment.  The monolithic design of the cavity, with the shear piezos, fibers, and alignment ferrules epoxied to a single block of aluminum (figure \ref{fig:apparatus}), should result in a mechanically stable cavity.  However, the membrane in the current setup is mounted 50~mm above a standard five-axis optical stage.  This relatively long and floppy mechanical connection between the cavity and the membrane is likely responsible for the small ripples in cavity resonance observed in figure 2\ref{fig:bandstructure}c and the detuning fluctuations we observe in the locked data.  This hypothesis is consistent with our observation of decreased cavity resonance frequency noise when the membrane is located at a node of the electric field (e.g., at about -110~nm in figure 2\ref{fig:bandstructure}a).
    
  Relative to free-space devices, it should be comparatively straightforward to integrate this type of fiber-coupled device with a cryogenic apparatus.  For example, optical fibers can readily be fed into windowless cryostats.  For cryogenic operation, the membrane would need to be mounted so that thermal contraction will not lead to large displacements of the membrane relative to the cavity.  This could be achieved by epoxying the membrane chip to the face of one of the glass ferrules.  This should also significantly improve the mechanical stability of the setup.
  
  A fiber-cavity also typically has a shorter cavity length and smaller beam waist than a free-space optomechanical cavity \cite{Thompson2008, Sankey2010, WilsonRegal2009, Vitali2011} and so can be used to create an optomechanical device with improved parameters.  Specifically, the shorter cavity length increases the optomechanical coupling $A$, corresponding to an increased radiation pressure per intra-cavity photon.  This increased coupling is particularly important when considering single-photon effects or when there are limits to the number of photons in the cavity due to bistability, heating, or power-handling.  The reduced beam waist allows the use of smaller, lighter membranes.  A smaller membrane is preferable because it can experience a reduced thermal Langevin force $S_{\text{F}}^{\text{thermal}}\propto m$ and can have a larger zero-point motion and fundamental resonance frequency.
  
  We anticipate that these improvements will result in a system where radiation pressure shot noise can be observed and will be within an order of magnitude of the 300~K thermal Langevin force.  In our current system, the maximum incident power is limited to about $\mathrm{8~\mu W}$ by the laser source.  There is no indication that this represents a fundamental limit, and the maximum incident power could be increased by more than an order of magnitude in future systems by improving the splices in our fiber setup.  At the current maximum power of $\mathrm{8~\mu W}$, 	$S_{\text{F}}^{\text{RPSN}}/S_{\text{F}}^{\text{thermal}}\approx10^{-4}$ which makes the detection of RPSN difficult.  However, we have observed membranes that, compared to the current membrane, have an order of magnitude smaller $m$ and two orders of magnitude smaller $\gamma_{\text{m}}$.  If such a membrane were to be incorporated into a fiber-cavity with the parameters reported above, then the system would have $S_{\text{F}}^{\text{RPSN}}/S_{\text{F}}^{\text{thermal}}>0.1$.
 
  In summary, we have demonstrated a passively-aligned high-finesse fiber-cavity with a membrane in the middle of the cavity.  The fiber-cavity is not only inherently fiber-coupled, but its small size also allows the use of smaller, lighter membranes and increases the linear and quadratic optomechanical coupling.  We have measured the expected optomechanical contributions to the mechanical resonance frequency and damping, and anticipate that this type of device will be able to detect radiation pressure shot noise at room temperature.
  
  This work has been supported by the DARPA/MTO ORCHID program through a grant from AFOSR.

\end{document}